\documentclass[10pt,aps,superscriptaddress,footinbib]{revtex4}

\usepackage[T1]{fontenc}
\usepackage[latin9]{inputenc}
\usepackage[letterpaper]{geometry}
\usepackage{float}
\usepackage{ae} 
\usepackage{color}
\usepackage[usenames,dvipsnames]{xcolor}

\usepackage{morefloats}
\usepackage{graphicx}
\usepackage{dcolumn}

\usepackage{amsmath,amssymb}
\usepackage{fmtcount}
\usepackage{upgreek}

\usepackage[english]{babel}

\input{epsf}
\input{epsfx}

\draft % marks overfull lines with a black rule on the right

\begin{document}

% Use the \preprint command to place your local institutional report number 
% on the title page in preprint mode.
% Multiple \preprint commands are allowed.
%\preprint{}

\title{Coherent spin control of a nanocavity-enhanced qubit in diamond} %Title of paper

% repeat the \author .. \affiliation  etc. as needed
% \email, \thanks, \homepage, \altaffiliation all apply to the current author.
% Explanatory text should go in the []'s, 
% actual e-mail address or url should go in the {}'s for \email and \homepage.
% Please use the appropriate macro for the type of information

% \affiliation command applies to all authors since the last \affiliation command. 
% The \affiliation command should follow the other information.

\author{Luozhou Li}
\altaffiliation{These authors contributed equally.}

\author{Tim Schr\"{o}der} 
\altaffiliation{These authors contributed equally.}

\author{Edward H. Chen} 
\altaffiliation{These authors contributed equally.}

\author{Michael Walsh} 

\author{Igal Bayn} 

\author{Jordan Goldstein} 

\author{Ophir Gaathon} 
\thanks{Currently at Diamond Nanotechnologies Inc. Boston, MA 02134, USA}

\author{Matthew E. Trusheim} 
\affiliation{Department of Electrical Engineering and Computer Science, Massachusetts Institute of Technology, Cambridge, MA 02139, USA}

\author{Ming Lu} 
 \affiliation{Center for Functional Nanomaterials, Brookhaven National Laboratory, Upton, NY 11973, USA}

\author{Jacob Mower} 
\affiliation{Department of Electrical Engineering and Computer Science, Massachusetts Institute of Technology, Cambridge, MA 02139, USA}

\author{Mircea Cotlet} 
 \affiliation{Center for Functional Nanomaterials, Brookhaven National Laboratory, Upton, NY 11973, USA}

\author{Matthew L. Markham} 

\author{Daniel J. Twitchen} 
\affiliation{Element Six, 3901 Burton Drive, Santa Clara, CA 95054, USA}

\author{Dirk Englund}
\email{englund@mit.edu}
\affiliation{Department of Electrical Engineering and Computer Science, Massachusetts Institute of Technology, Cambridge, MA 02139, USA}

%\author{XXX}
%\affiliation{Columbia}

% Collaboration name, if desired (requires use of superscriptaddress option in \documentclass). 
% \noaffiliation is required (may also be used with the \author command).
%\collaboration{}
%\noaffiliation

\newcommand{\QnvA}{1550}
\newcommand{\QnvAerror}{100}
\newcommand{\TnvA}{230~$\upmu$s}

% NV-A, Ey
\newcommand{\FnvAex}{12}
\newcommand{\FnvAexerror}{1}
\newcommand{\BETAnvA}{26}
\newcommand{\BETAnvAerror}{1}

% NV-A, Ex
\newcommand{\FnvAey}{10}
\newcommand{\FnvAeyerror}{1}
\newcommand{\BETAnvAey}{23}
\newcommand{\BETAnvAeyerror}{1}

% NV-B
\newcommand{\FnvB}{60}
\newcommand{\FnvBerror}{12}
\newcommand{\BETAnvB}{82}
\newcommand{\BETAnvBerror}{4}
\newcommand{\LifetimeBenhanced}{6.7}
\newcommand{\LifetimeBinhibited}{18.4}
\newcommand{\LifetimeBenhancedError}{0.1}
\newcommand{\LifetimeBinhibitedError}{0.5}
\newcommand{\CnvB}{2.45}
\newcommand{\CnvBerror}{1}
\newcommand{\QnvB}{3700}
\newcommand{\QnvBerror}{100}
\renewcommand{\thefootnote}{\alph{footnote}}

\date{\today}

\begin{abstract}
A central aim of quantum information processing is the efficient entanglement of multiple stationary quantum memories via photons. Among solid-state systems, the nitrogen-vacancy (NV) centre in diamond has emerged as an excellent optically addressable memory with second-scale electron spin coherence times. Recently, quantum entanglement and teleportation have been shown between two NV-memories, but scaling to larger networks requires more efficient spin-photon interfaces such as optical resonators. Here, we demonstrate such NV-nanocavity systems with optical quality factors approaching 10,000 and electron spin coherence times exceeding 200~$\upmu$s using a silicon hard-mask fabrication process. This spin-photon interface is integrated with on-chip microwave striplines for coherent spin control, providing an efficient quantum memory for quantum networks.
\end{abstract}

% insert suggested PACS numbers in braces on next line

\maketitle %\maketitle must follow title, authors, abstract and \pacs

% Body of paper goes here. Use proper sectioning commands. 
% References should be done using the \cite, \ref, and \label commands
\begin{figure} 
	\centering
		\includegraphics[width=\textwidth]{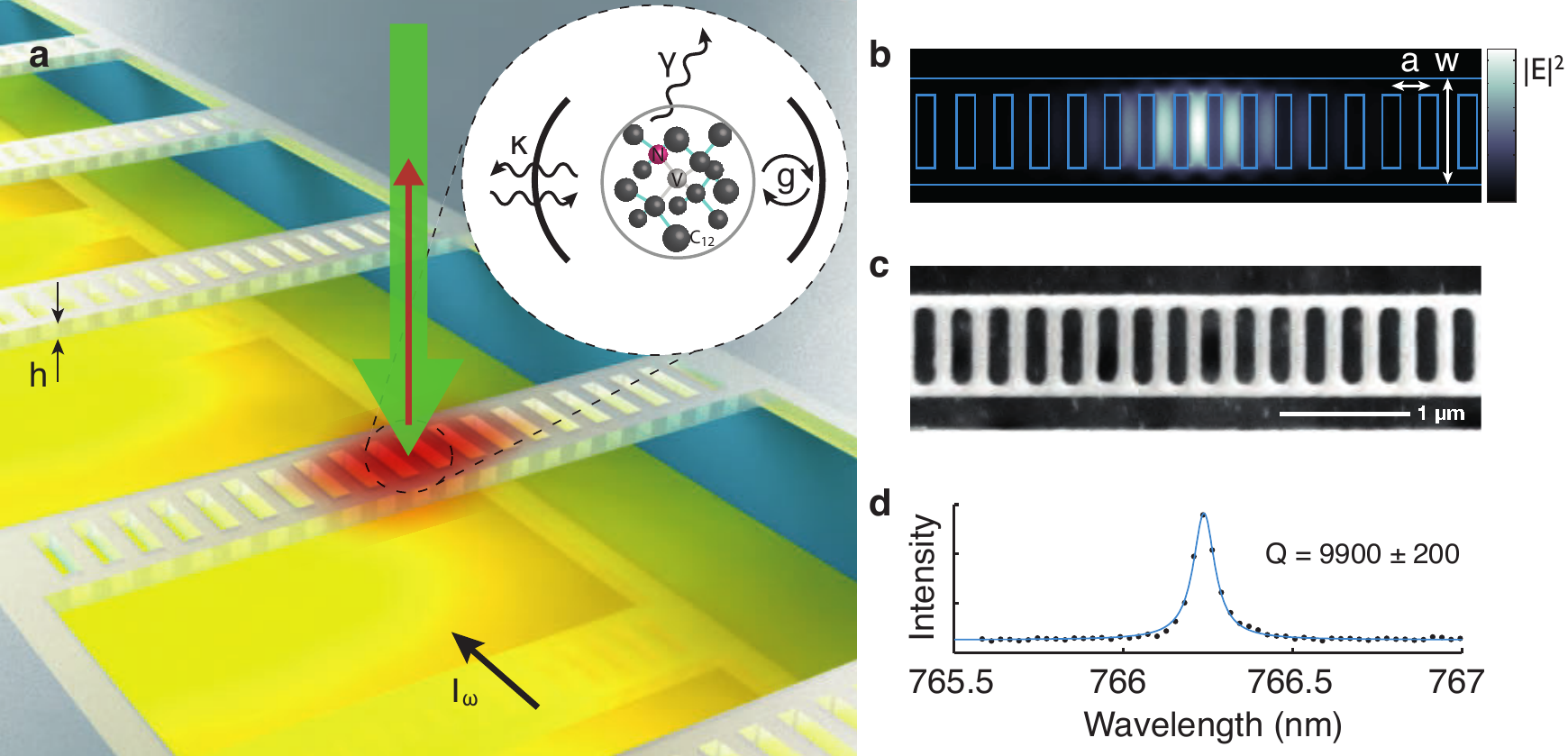}
	\caption{\footnotesize On-chip NV-nanocavity system in diamond.
	% Figure legends: Figure legends for Articles or Letters begin with a brief title for the whole figure and continue with a short description of each panel and the symbols used, focusing on describing what is shown in the figure and de-emphasizing methodological details. The meaning of all error bars and how they were calculated should be described. Each legend should total no more than 250 words.
	a, The diamond PhC cavities are integrated on a Si substrate with metallic striplines for coherent spin control and optically addressed using a confocal setup with 532~nm CW excitation and photoluminescence collected $>630$~nm. The inset shows the NV-nanocavity system with $g$ the NV-nanocavity Rabi frequency, $\gamma$ the NV natural spontaneous emission (SE) decay rate, and $\kappa$ the cavity intensity decay rate. The NV consists of a substitutional nitrogen atom adjacent to a vacancy in the diamond lattice. $I_\omega$ denotes the current through the stripline, and $h$ the PhC thickness.
	%Illustration of PhC diamond cavities containing NVs integrated onto a Si substrate with metallic striplines and optically addressed by a confocal setup. Coherent control of the electron spin of the NV ground state is achieved via the microwave stripline. The inset illustrates the atomic structure of an NV and describes the relevant parameters of our NV-cavity system. 
	b, Simulated electric field intensity for the optimised fundamental cavity mode. The PhC has a width $W$ and a lattice constant varying from $0.9a$ at the centre to $a=220$~nm over five periods.
	c, Scanning electron micrograph (SEM) of a representative cavity structure.
	d, Measured cavity resonance (dots) with a quality factor $Q\sim9,900\pm200$ from a Lorentzian fit (blue line).
	}
\label{fig: 1}
\end{figure}

The coupling between photons and quantum states of an emitter is efficient in the strong Purcell regime, in which the emitter interacts primarily with one optical mode. This regime is reached when the overall Purcell enhancement exceeds one ($F > 1$, see Fig.~1a)\cite{2008.OpEx.Hollenberg.NV_cavity}. When the NV zero-phonon line (ZPL) is coupled to a cavity with quality factor $Q$ and mode volume $V_{mode}$, the spectrally-resolved SE rate is enhanced by the Purcell factor
\begin{equation}
F_{ZPL} = \xi F_{ZPL}^{max}\frac{1}{1+4Q^2(\lambda_{ZPL}/\lambda_{cav}-1)^2} \\
\end{equation}
%\begin{equation}
%F_{ZPL}^{max}=\frac{3}{4\pi^2}\left(\frac{\lambda}{n}\right)^3\frac{Q}{V_{mode}} \\
%\end{equation}
%\begin{equation}
%\xi=\left(\frac{\vert\vec{\mu}\cdot \vec{E}\vert}{\vert\vec{\mu}\vert\vert \vec{E}_{max}\vert}\right)^2 \\
%\end{equation}
\noindent where $F_{ZPL}^{max}=\frac{3}{4\pi^2}\left(\frac{\lambda_{cav}}{n}\right)^3\frac{Q}{V_{mode}}$ is the maximum  spectrally-resolved SE rate enhancement and $\xi=\left(\frac{\vert\vec{\mu}\cdot \vec{E}\vert}{\vert\vec{\mu}\vert\vert \vec{E}_{max}\vert}\right)^2$ quantifies the angular and spatial overlap between the dipole moment ($\vec{\mu}$) and the cavity mode electric field~($\vec{E}$)\cite{santori_single-photon_2010}. The highest $F_{ZPL}^{max}$ can be realised in photonic crystal (PhC) nanocavities due to their small mode volumes\cite{tiecke_nanophotonic_2014}, $V_{mode}\sim(\lambda/n)^3$. 1D and 2D PhC cavities in diamond\cite{riedrich2011one, Faraon_coupling_2012,hausmann_coupling_2013} have reached $Q$ factors of 6,000 and 3,000, and $F_{ZPL}$ up to 7 and 70, respectively. But thus far, the longest spin coherence time of cavity-enhanced NV centres has been less than 1~$\upmu$s, limiting their suitability as a quantum memory\cite{englund_deterministic_2010}. Here, we considered a new fabrication process to produce NV-nanocavity systems in the strong Purcell regime with long spin coherence times of cavity-coupled NVs and greatly improved cavity $Q$ factors. 
%$F_{ZPL}^{max}$ can be especially high in photonic crystal nanocavities with volumes about $(\lambda/n)^3$\cite{ring cavity,2D,becher,loncar}. Previous work on cavities has shown $Q$ values in diamond up to 6,000 and spectrally resolved Purcell enhancement of the zero-phonon-line (ZPL) transitions with the highest $F_{ZPL}$ of $70$\cite{riedrich2011one,faraon2012coupling, hausmann2013coupling}, but thus far, there has been no demonstration of coherent control over long-lived NVs ($T_2\gg10\upmu$s). 

%Here, we introduce a new fabrication process to produce diamond nanocavities with greatly improved $Q$ factors. These NV-nanocavity systems have long spin coherence times, and we also demonstrate a cavity-coupled NV in the strong Purcell regime. 

%\section{Results}
%\subsection{Simulation}
The cavities were designed using finite-difference time-domain (FDTD) simulations\cite{eichenfield2009picogram} to maximise $F_{ZPL}^{max}$ by optimising the ratio of $Q/V_{mode}$ (Supplementary Section 1).
%The cavities were designed by finite-difference time-domain (FDTD) simulations\cite{gong2010photonic,eichenfield2009picogram} to maximise the overall collection efficiency of coherent ZPL radiation through the cavity mode by (i) a high $Q/V_{mode}$ ratio and (ii) high out-coupling efficiency due to low effective refractive index in the cavity centre\cite{tsai2011directional}. 
As shown in Fig.~1, the nanocavity is based on a suspended one-dimensional diamond PhC structure with lattice constant $a$, beam width $w=2.4a$, and thickness $h=0.7a$. A linear increase of the lattice constant from $0.9a$ to $a$ in increments of $0.02a$ per period away from the centre defines the cavity defect state. The fundamental cavity mode of the optimised structure yielded $Q=6.02\times10^5$ and $V_{mode}=1.05(\lambda/n)^3$. 

\begin{figure} 
	\centering
		\includegraphics[width=85mm]
		{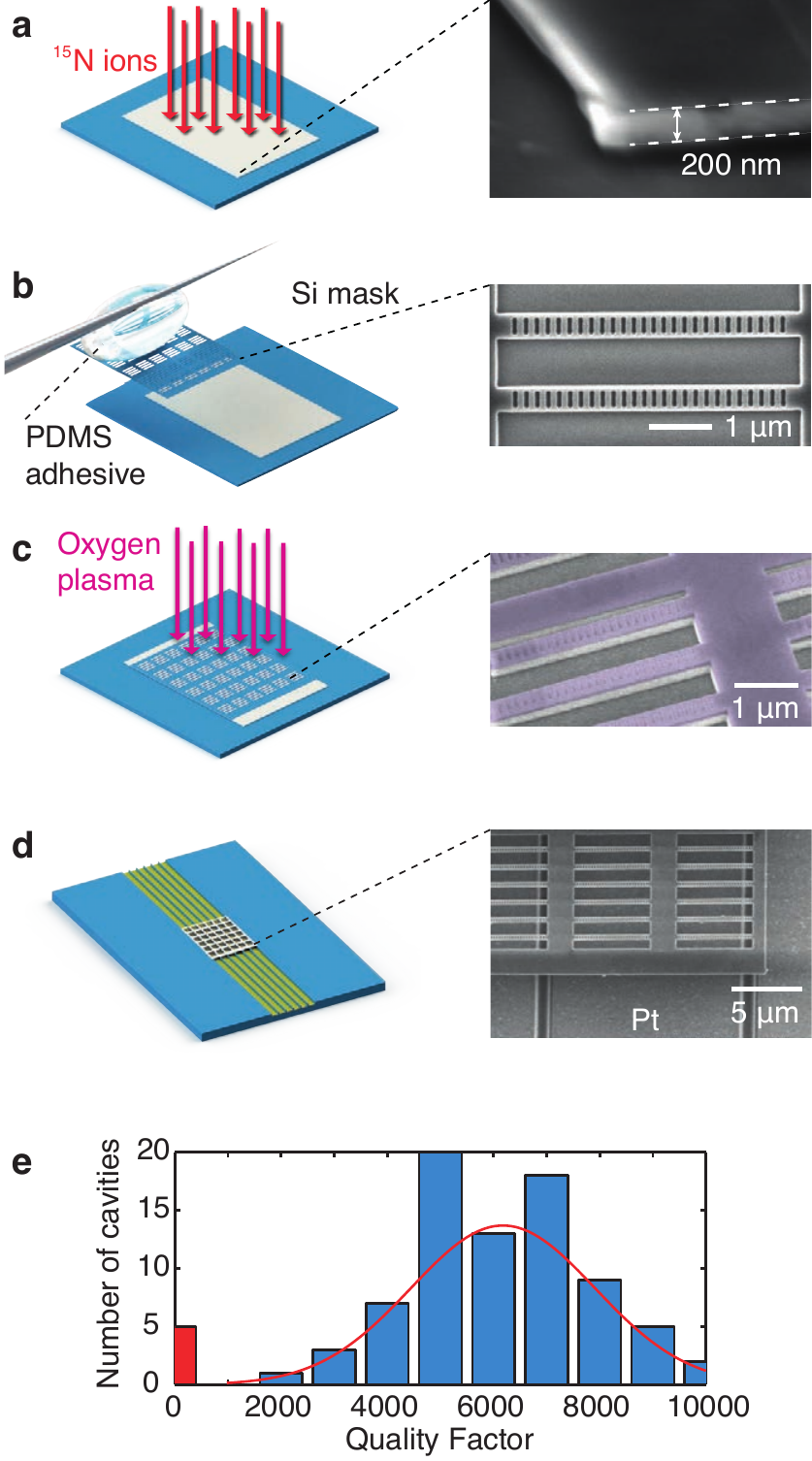}
	\caption{\footnotesize Fabrication procedure (left column) and SEM of representative structures (right column). 
%A 5~$\upmu$m diamond membrane on a Si substrate was thinned by reactive ion etching (RIE) to $\sim$200~nm thickness.
	a, NVs were created $\sim$100~nm below the surface of the diamond membranes by implantation of $^{15}$N atoms and subsequent annealing at 850$^\circ$C. \textit{Right}: SEM of 200~nm membrane. 
	b, Si masks were patterned on SOI, released, and transferred onto diamond membranes. \textit{Right}: Patterned Si mask before transfer.
	c, Oxygen RIE was used to pattern diamond membranes. \textit{Right}: The false-colour SEM shows the Si mask (purple) on diamond after oxygen etching. 
	d, Patterned diamond membrane on microwave striplines for optical and spin characterisation. \textit{Right}: SEM of diamond PhC structures above metallic striplines in Si channels. 
	e, Distribution of cavity $Q$ factors from one fabrication run. 78 (blue bars) of 83 cavities showed resonances in the range of 600-770~nm, while five (red bar) showed no resonances in this wavelength range. The mean $Q$ is 6,200.
	}
\label{fig: 2}
\end{figure}

%-- the most studied semiconductor material in microelectronics and photonics --
%\subsection{Nanofabrication using silicon masks}
The cavities were patterned in high-purity single-crystal diamond using a new fabrication process that employs silicon (Si) membranes as etch masks. The diamond was fabricated by microwave plasma assisted chemical vapour deposition (CVD), polished to 5 $\upmu$m thickness, and finally thinned to $\sim$200~nm using a combination of chlorine and oxygen reactive ion etching (Methods). NVs were created by implantation of $^{15}$N and subsequent annealing (Fig.~2a). The Si masks were produced by electron beam lithography and cryogenic plasma etching (sulfur hexafluoride and oxygen) from silicon-on-insulator wafers with $\sim$220~nm-thick device layers\cite{lipson_silicon_2007}. This resulted in high-quality masks approximately $100\times100~\upmu\textnormal{m}^2$ in area. These were subsequently placed onto the diamond membranes using a transfer process described in the Methods (Fig.~2b). This Si mask transfer process enables nano-patterning without the need for spin-coating resist onto substrates and is compatible with samples sizes down to several tens of square micrometers. We used oxygen plasma\cite{li2013reactive} to etch the Si mask pattern into the pre-thinned $\sim$200~nm diamond membranes (Fig.~2c). After mask removal, the patterned diamond membranes were transferred onto a Si chip with integrated microwave striplines (Fig.~2d). Because the silicon mask can be fabricated with excellent quality, thanks to the availability of mature fabrication technology for this material, this process yields diamond PhCs with low surface roughness and uniform, vertical sidewalls. We observed a high yield (94~$\%$) of cavities with resonances close to NV ZPL in a single fabrication run, with a mean $Q$ of 6,200 and a maximum $Q$ of $9,900\pm200$~(Fig.~2e). Cavity resonances spectrally lower than 637 nm are suitable for NV ZPL coupling while longer wavelength resonances can be blue-detuned by thermal oxidation and oxygen plasma etching\cite{riedrich2011one, hausmann_coupling_2013}.  

\begin{figure} 
	\centering
		\includegraphics[width=150mm]{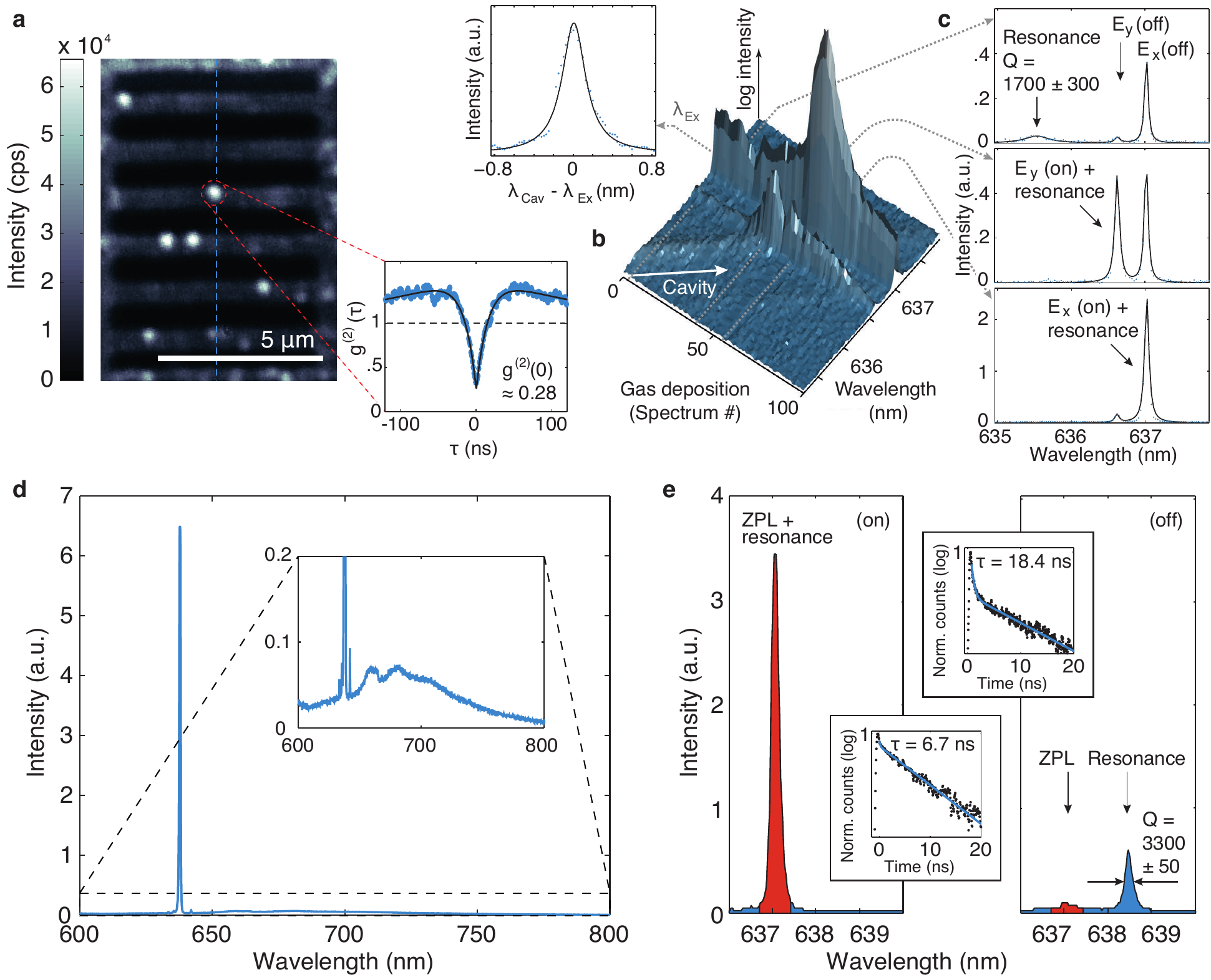}
	\caption{\footnotesize Optical characterisation of NV-nanocavity systems.
	a, PL confocal image of diamond PhC structures. Single NVs are identified by circular white spots. System A: The dotted red circle shows a single NV close to the cavity centre (indicated by the blue dotted line). Inset: Normalised second-order auto-correlation measurement with $g^{(2)}(0)=0.28$.  
	 b, Gas tuning of System A. The logarithmic plot shows the cavity resonance and two strain-split ZPL branches from a single NV ($E_y$ and $E_x$, $2\Delta=286$~GHz, see Fig.~4a). As the gas condensation red-shifts the cavity resonance, it sequentially enhances the two ZPL branches. The inset shows the intensity of the $E_x$ ZPL transition as a function of cavity detuning. This curve follows the expected Lorentzian dependence of the Purcell enhancement given by Eqn.~1, and shows that the cavity Q factor remains constant throughout the tuning process.~%Right inset: 
	 c, Spectra of system A in the uncoupled (top) and coupled cases with $\lambda_{cav}=\lambda_{E_y}$  (centre) and $\lambda_{cav}=\lambda_{E_x}$ (bottom, note the difference in scaling). The black lines are Lorentzian fits to the data, yielding $Q=1,700\pm300$ for the cavity.
	 d, System B at maximum Purcell enhancement. The inset shows a close-up of the spectrum. The ZPL transitions of four individual NVs (including the cavity-coupled ZPL) are visible, each with a different strain-induced spectral position. The accumulated PSBs of these NVs are also apparent.  
	 e, High resolution spectra of system B of the coupled (left) and uncoupled (right) case. The insets show the lifetime measurements corresponding to $\tau_{on}=\LifetimeBenhanced$~ns and $\tau_{off}=\LifetimeBinhibited$~ns. %The black lines are Lorentzian fits to the data.
	 }  
	 \label{fig: 3}
\end{figure}

%\subsection{Optical characterisation}
We optically characterised samples at ambient and cryogenic ($\sim$18~K) temperatures using homebuilt confocal microscope setups with 532~nm continuous-wave (CW) laser excitation. Photoluminescence (PL) imaging (Fig.~3a) was used to identify NVs spatially within cavity centres, and spectral measurements determined the separation between the NV ZPL transitions and cavity resonances. Cavity system A (circled in Fig.~3a) contains a single NV, as verified by antibunching in the second-order auto-correlation function. Fig.~3c plots the initial PL spectrum of system A, showing a cavity peak ($Q=1,700\pm300$) blue-detuned from the ZPL, as well as two ZPL branches $E_x$ and $E_y$. These are split by 286~GHz due to local strain in the diamond lattice\cite{batalov_low_2009}. As shown in Fig.~3b, the cavity resonance was then gradually red-shifted by gas deposition\cite{srinivasan_optical_2007} to overlap with the NV ZPL transitions, resulting in strong PL enhancements (Methods). A rate equation model is used to analyse the transition dynamics of the NV centre to determine $F_{ZPL}$ (Supplementary Section 2). In this simplified five-level model, we account for the SE rates of the PSB transitions and the different ZPL transition rates for the on- and off-resonance cases. The zero-phonon excited state to ground state transition rates of the NV are assumed to be enhanced by the factor $F_{ZPL}$. One important input parameter of our model is the fraction of ZPL to total intensity, quantified by the Debye-Waller factor\cite{zhao2012suppression}, which we estimate from an off-resonance spectrum to be $\textit{DW}=0.028$. By substituting this factor into the rate equation model, we determine $F_{ZPL}$ of $8$~($15$) for transition E$_x$~(E$_y$) in System A. To show that our simplified rate equation model gives a reliable prediction of $F_{ZPL}$, we considered a second analysis method. By comparing the ZPL intensity for coupled and uncoupled cases (Fig.~3c), we can determine the ZPL SE coupling efficiency into the cavity mode, $\beta = I_{ZPL}^{cavity}/(I_{ZPL}^{cavity}+I_{PSB})$. This method is valid in the weak excitation limit where the low population of the excited state does not influence the ratio of $I_{total}^{on}/I_{total}^{off}$. From $\beta\sim F/(F+1)$, we can then deduce the increase in SE rate $F_{ZPL}= F/\textit{DW}=10~(17)$ for the E$_x$ (E$_y$) transition, yielding similar values compared to the rate equation analysis.

Since the Purcell enhancement depends strongly on the spatial and angular overlap, $F_{ZPL}$ is generally much lower than the maximum possible value, $F_{ZPL}^{max}$, especially in samples with low NV density ($\sim$$1~\textnormal{NV}/\upmu$m$^{2}$ in the case of System A). Moreover, for the \{100\} diamond crystal used here, the maximum Purcell factor is reduced to $F_{ZPL}^{max*}=\cos^2({35.3^\circ})F_{ZPL}^{max}$, since $35.3^\circ$ is the smallest angle between the transverse-electric (TE) cavity field and the NV dipole (i.e., crystal) orientation. Using the rate equation model for System A, we calculate an overlap factor $\xi=F_{ZPL}/F_{ZPL}^{max*}=0.1~(0.18)$ for transition $E_x$~($E_y$). The difference in $\xi$ is attributed to the different orientations of the two orthogonal NV dipoles with respect to the TE-cavity mode\cite{doherty2013nitrogen}. 

To investigate NV-nanocavity systems in the strong Purcell regime, we studied another sample with the same cavity designs and a higher density of NVs ($\sim$$10/\upmu$m$^2$). Fig.~3d shows the PL spectrum of NV-nanocavity system B with $Q_B=3,300\pm50$. Out of four ZPL transitions, one was strongly enhanced by the cavity mode; we attribute the remaining ZPL transitions to spatially decoupled NV centres within the $\sim$$2~\upmu$m diameter microscope collection spot through the cryostat window. We observed both a change in spontaneous emission lifetime from $\tau_{off}\sim\LifetimeBinhibited$~ns to $\tau_{on}\sim\LifetimeBenhanced$~ns and a strong increase in emission from this NV ZPL when tuned onto resonance with the cavity~(Fig.~3d,e). Due to the presence of multiple ZPL transitions and their accumulated PSBs, we cannot measure directly their individual $DW$ factors, which are required to precisely determine the SE rate enhancement $F_{ZPL}$. We therefore used two independent measurements to determine $DW$ and $F_{ZPL}$: (i) the rate equation model (as done for system A), and (ii) the radiative lifetime modification according to $F_{ZPL} = (\tau_{bulk}/\tau_{on}-\tau_{bulk}/\tau_{off})/DW$ (Fig.~3e). Solving this system of equations gives $DW=0.019$ and $F_{ZPL}=62$ for a measured $\tau_{bulk}\sim 12.5$~ns. We calculate $\beta=0.54$ and an overall Purcell factor $F=1.2>1$, indicating that System B is in the strong Purcell regime (Supplementary, Section 2). The $DW$ has been reported in a range from 0.01 to 0.19\cite{zhao2012suppression}; because of this wide variability, we emphasize that it is important to obtain the DW factor from separate measurements, as this strongly influences the estimated value of $F_{ZPL}$. 

%The overlap factor ($\xi=0.36$) shows better mode overlap with the cavity mode than in System A as expected from the much higher SE rate enhancement.

%By again comparing the coupled and uncoupled ZPL cases with $\textit{DW}$ of 0.026 (Fig.~3e), we estimate a $\beta$-factor of $0.71\pm0.03$ which corresponds to a very strong SE rate enhancement of $F_{ZPL}=\FnvB\pm\FnvBerror$, closer to the highest possible value of $F_{ZPL}^{max*}\sim159$. Direct lifetime measurements yielded a change from $\tau_{off}\sim\LifetimeBinhibited\pm\LifetimeBinhibitedError$~ns to $\tau_{on}\sim\LifetimeBenhanced\pm\LifetimeBenhancedError$~ns, which supports Purcell enhancement obtained from $\beta$-factor (Supplementary Section 2).
%and deduce a spontaneous emission lifetime\cite{Faraon_coupling_2012} of $\tau_{est}\sim2.3\pm$0.3~ns.

%Direct lifetime measurements yielded $\tau_{on}\sim\LifetimeBenhanced\pm??$ and $\tau_{off}\sim\LifetimeBinhibited\pm??$ for the  coupled and uncoupled case. This gives a lower bound for the spectrally resolved Purcell enhancement of $F\sim50?$. We attribute the difference to degradation of the NV-cavity system by extended laser exposure and thermal treatment at $475^{\circ}C$ (See Supplementary$\textit{We have to add this info to the Suppl.}$) before performing the lifetime measurements.

%Slightly after the point of maximum spectral overlap, the change in spontaneous emission lifetime was measured to be \LifetimeBenhanced$\pm??$~while for a far detuned cavity the inhibited spontaneous emission lifetime was \LifetimeBinhibited$\pm??$. 

\begin{figure} 
	\centering
		\includegraphics[width=\textwidth]{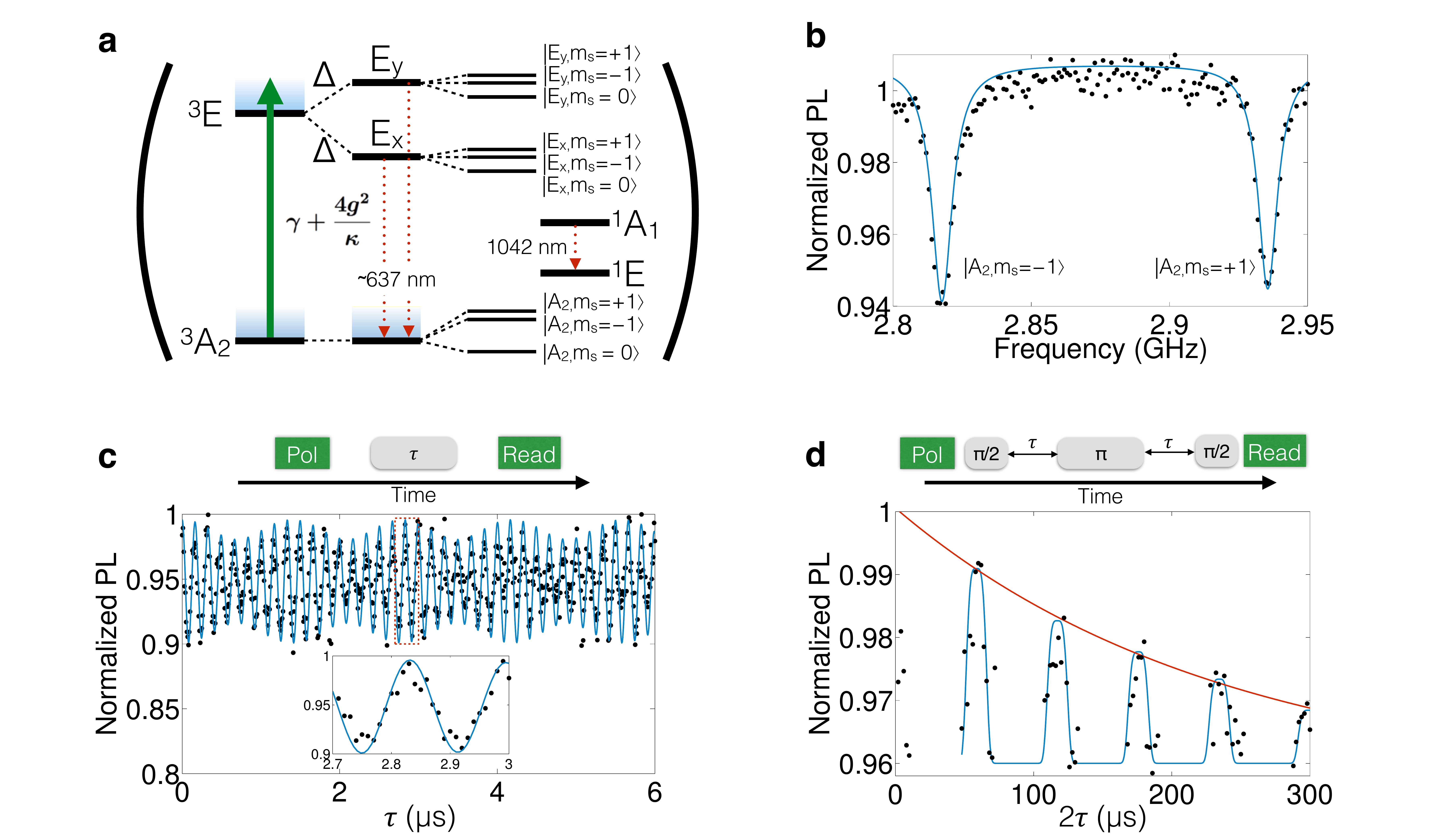}
	\caption{\footnotesize Electronic structure and coherent spin control. a, Energy level diagram of an NV centre in a nanocavity. The two excited state sub-levels, $E_x$ and $E_y$, are split by $2\Delta\sim$~286~GHz due to strain in the lattice for NV system A (circled in Fig 3a). The effective SE rates of these excited states are modified by the nanocavity from $\gamma$ to $\gamma+\frac{4g^2}{\kappa}$. For both the excited and ground state triplets, the $\left|m_s=0\right>$ and $\left|m_s=\pm1\right>$ states are split by the crystal field splitting, and the degeneracy of the $\left|m_s=\pm1\right>$ states is lifted by an applied magnetic field. 
	b, The optically detected magnetic resonance spectrum of the ground state triplet in NV System A with an applied field of $\sim$2~mT. 
	c, By addressing only the transition between the $\left|m_s=0\right>$ and $\left|m_s=-1\right>$ states, we can coherently drive Rabi oscillations without observing any decay of the envelope function which indicates a $T_2'$$~\textgreater ~6~\upmu$s. The Rabi oscillations contain two frequency components which we attribute to a hyperfine interaction with a nuclear spin. The inset above shows the Rabi sequence of polarisation, pulsed MW excitation, and optical read-out. %The beating pattern is due to the hyperfine interaction with the {\color{blue}{XXX}} by 6~MHz and 6.7~MHz. 
	d, Using the $\pi/2$ and $\pi$ pulse durations from the Rabi measurement, we then applied a spin-echo sequence, determining the $T_2$ to be \TnvA. The inset above shows the Echo sequence of polarisation, pulsed $\pi/2$ and $\pi$ MW pulses, and optical read-out.}
\label{fig: 4}
\end{figure}

%\subsection{Spin measurements}
The cavity-coupled NV centres exhibit excellent spin coherence times $(T_2)$ similar to the parent CVD crystal. Fig.~4a shows the spin-preserving $E_x$ and $E_y$ optical transitions\cite{doherty2013nitrogen} of NV-nanocavity System A with associated magnetic sublevels $m_s=-1,0,1$. A small static magnetic field of $\sim$2~mT was applied along the NV axis to lift the degeneracy of the $m_s=\pm 1$ spin states. The metastable spin singlet states ($E_1$,$A_1$) enable optically detected magnetic resonance (ODMR) measurements\cite{doherty2013nitrogen}, but we do not expect the transition rate between these levels to be modified by the cavity because the PhC bandgap, targeted near the ZPL at 637 nm, does not overlap with the 1042~nm line. Fig.~4b shows ODMR under continuous optical and MW excitation, indicating the $m_{s}=0\rightarrow \pm 1$ spin transitions. Separate Rabi oscillation measurements (Fig.~4c) under pulsed excitation indicate an ensemble phase coherence time\cite{hanson_room-temperature_2006}  of $T_2'>6~\upmu$s. %The Rabi oscillations contained two frequency components which we attribute to a hyperfine interaction with a nuclear spin. 
The phase coherence time $T_2$ is measured using a Hahn echo to cancel the dephasing by quasi-static magnetic fields\cite{hanson_room-temperature_2006}. From the single-exponential\cite{hodges_long-lived_2012} decay envelope of the revivals in Fig.~4d, we estimate $T_{2}\sim$~\TnvA. Such $T_2$ values are typical for the parent diamond crystal, indicating that our nanofabrication process preserves long electron spin coherence times. % nanopillar paper with long T2
This coherence time is more than two orders of magnitude longer than previously reported values for cavity-coupled NV centres\cite{englund_deterministic_2010} and semiconductor quantum dots\cite{carter2013quantum, tartakovskii_quantum_2014}.%

We use the relevant NV-nanocavity parameters to determine the possible impact of our system on established and potential future applications. NV-nanocavity system~B lies in the strong Purcell regime %with rates of $(\kappa, g_{ZPL}, \gamma_{ZPL})/(2\pi) = $(110$\pm$6~GHz, 15$\pm$7~GHz, 50$\pm$50~MHz) 
with $\beta= 0.54$, which would lead to a $\sim$800-fold increase in entanglement generation rates between two distant NVs compared to present schemes without cavity enhancement\cite{bernien2013heralded} (Supplementary Section 3). Recently achieved quantum teleportation rates would also significantly benefit from this speed up\cite{pfaff2014unconditional}. 
%Transition E$_y$ of NV system A has $(\kappa, g_{ZPL}, \gamma_{ZPL})/(2\pi) = (304\pm15~\textnormal{GHz}, 7\pm3~\textnormal{GHz}, 50\pm50~\textnormal{MHz})$ and E$_x$ has $(\kappa, g_{ZPL}, \gamma_{ZPL})/(2\pi) = (303\pm15~\textnormal{GHz}, 6\pm3~\textnormal{GHz}$). 
%In this regime, a $\beta$-factor of $\BETAnvA\pm\BETAnvAerror$, leads to a $\sim$44-fold increase in entanglement generation rate  An even further increase could be achieved using
These estimations indicate that coupling long-lived NVs to single-crystal diamond cavities is a critical step towards long-distance quantum entanglement and large-scale quantum networks. Furthermore, such cavity-coupled NV systems would enable implementations involving non-linear interactions for quantum memories\cite{heshami2013raman} and quantum repeaters\cite{childress2006fault}.

%This one-dimensional cavity design can be efficiently coupled to waveguides within an integrated network\cite{quan2010photonic}. [[PUT IN somewhere?]]

%NV-coupled diamond PhC nanocavities with consistently high $Q$ factors, long spin-coherence times, and high NV cavity-mode overlap. 

In conclusion, we have introduced a fabrication process for the creation of NV-nanocavity systems in the strong Purcell regime with consistently high Q factors while preserving the long spin coherence times of NVs\cite{bar2013solid}. These systems enable coherent spin control of cavity-coupled semiconductor qubits with coherence times exceeding $200~\upmu$s --- an increase by two orders of magnitude over previous cavity-coupled solid-state qubits\cite{englund_deterministic_2010, carter2013quantum, tartakovskii_quantum_2014}. Our on-chip architecture could be used to efficiently scale NV-nanocavity systems to many quantum memories connected via photons\cite{CZKM1997PRL,kimble_quantum_2008,kim2013quantum,noda2007spontaneous}. The membrane-transfer process introduced here is well-suited for building such networks as it allows the screening and subsequent integration of high-performance NV-nanocavity systems\cite{wolters_deterministic_2014,burkard_cavity-mediated_2014} into photonic integrated circuits equipped with microwave circuits for multiple electron and nuclear spin control\cite{neumann2010single,dolde2013room}, waveguide-integrated superconducting detectors, and low-latency logic devices for feed-forward\cite{benson_assembly_2011}. Spatial implantation of NVs into the mode field maximum or cavity fabrication around a single NV\cite{riedrich-moller_deterministic_2014} appear promising to increase the NV-nanocavity overlap probability. Many of the schemes discussed above require coherent optical control of single or multiple NV spins in cavities that exhibit low spectral diffusion and lifetime-limited ZPL transitions; recent work on near-surface implanted NVs indicates that this should be possible\cite{wolters2013measurement, chu_coherent_2014}. With these advances, multiple NV-nanocavity systems operating in the strong Purcell regime and having long spin coherence times would form scalable quantum memories for quantum repeaters\cite{childress2006fault}, spin-based microprocessors\cite{awschalom_quantum_2013}, and quantum networks\cite{o2009photonic}.

\textbf{METHODS}

In the first fabrication step, high-purity ($^{14}$N $<100$ ppb) single-crystal diamond plates were grown by microwave-plasma assisted CVD and were laser cut to a thickness of $\sim$200~$\upmu$m.  The plates were polished down to $\sim$5~$\upmu$m membranes using a cast iron scaif. For the creation of NVs, a layer of nitrogen atoms was implanted at 80~keV energy and located $\sim$100~nm from the surface. System A was implanted at a dosage of 5$\times10^{10}$ $^{15}$N/cm$^2$ and System B at 5$\times10^{11}$ $^{15}$N/cm$^2$. The membrane was annealed in a MTI OTF-1500X-4 vacuum furnace (1.5$\times10^{-6}$~mbar) for two hours at 850$^\circ$C to mobilise lattice defects, which combine with $^{15}$N atoms to form NV centres. Next, the membrane was turned over and thinned down to $\sim$200~nm using plasma etching (Oxford ICP-RIE) with a mixture of chlorine and argon gases at a etch rate of $\sim$2~$\upmu$m/hour. This recipe yielded a smooth surface ($\textnormal{RMS}<1~\textnormal{nm}$) after 4.8~$\upmu$m etching. The thinned membranes generally exhibited inhomogeneous thicknesses (100~nm to~300 nm) over hundreds of micrometers. The membrane was divided into tens of smaller pieces, which were transferred onto separate silicon substrates using a polydimethylsiloxane (PDMS)-tipped tungsten probe. The Si PhC masks were designed and fabricated to match the thickness of each membrane so that cavity resonances would fall near the NV's ZPL, and then transferred onto the membranes using a PDMS-tipped probe. Oxygen plasma dry etching (Trion RIE at 20~sccm gas flow, 50~mTorr pressure and 100~W power) was used to transfer the pattern into the membranes. After the etch, little erosion was found on the silicon PhC masks. A tungsten probe was used to remove silicon masks from diamond membranes. Finally, an SF$_6$ isotropic dry etch removed the silicon underneath to suspend the cavity structures. MW striplines were produced separately on intrinsic silicon using a standard semiconductor fabrication process, followed by a lift-off step for metal deposition into the silicon trenches. Finally, the diamond devices were integrated into the MW architecture using a PDMS-tipped probe.

\noindent
Characterisation of the optical properties of the sample at cryogenic temperatures was performed via photoluminescence measurements in a continuous flow He cryostat (CCS-XG-M/204N, Janis) at $\sim$18~K. The sample was mounted inside the isolation vacuum and accessed through a window-corrected objective (LD Plan-Neofluar 63x, Zeiss NA = 0.75). The NVs contained in the diamond cavity structures were excited with a 532~nm continuous-wave (CW) laser (Coherent Compass 315M). Fluorescence from the sample was collected in a confocal configuration and sent to fibre-coupled single photon detectors (SPCM-AQR, Perkin Elmer), while spectra were taken via free-space coupling into a spectrometer (Isoplane SCT320, Princeton Instruments). To spectrally tune the cavity mode into resonance with the ZPL, the cryostat was equipped with a nozzle near the cold-finger for controlled gas flow onto the sample. This feature can be used for condensation and ice formation of gas (e.g. Xe) onto the sample, hence changing the effective refractive index of the diamond membrane. This refractive index change allows for spectrally red-tuning cavity resonances at a rate of $\sim$8~pm/s. To take full advantage of this tuning technique the cavities were designed to have resonances spectrally blue-shifted from the ZPL. Xe gas can then be used to achieve precise, spectrometer-controlled, in-situ tuning of the cavity to overlap its resonance with the ZPL. %Xenon gas was introduced into the cryostat through a nozzle directed at the sample and positioned $\sim$1 cm away from it. Spectra were then collected during the deposition process. As expected, the diamond Raman line and NV ZPL remain constant during the deposition while the cavity resonance red shifts as the effective index of the material is altered. 
We note that the cavity tuning was observed within seconds of the Xe being released indicating no further gas dynamics. Reheating the sample to room temperature reverses the tuning. Using this procedure, we were able to repeatedly tune over a range of $\sim$31~nm without significant degradation of the cavity $Q$.

% If in two-column mode, this environment will change to single-column format so that long equations can be displayed. 
% Use only when necessary.
%\begin{widetext}
%$$\mbox{put long equation here}$$
%\end{widetext}

% Figures should be put into the text as floats. 
% Use the graphics or graphicx packages (distributed with LaTeX2e).
% See the LaTeX Graphics Companion by Michel Goosens, Sebastian Rahtz, and Frank Mittelbach for examples. 
%
% Here is an example of the general form of a figure:
% Fill in the caption in the braces of the \caption{} command. 
% Put the label that you will use with \ref{} command in the braces of the \label{} command.
%
% \begin{figure}
% \includegraphics{}%
% \caption{\label{}}%
% \end{figure}

% Tables may be be put in the text as floats.
% Here is an example of the general form of a table:
% Fill in the caption in the braces of the \caption{} command. Put the label
% that you will use with \ref{} command in the braces of the \label{} command.
% Insert the column specifiers (l, r, c, d, etc.) in the empty braces of the
% \begin{tabular}{} command.
%
% \begin{table}
% \caption{\label{} }
% \begin{tabular}{}
% \end{tabular}
% \end{table}

% If you have acknowledgments, this puts in the proper section head.
%\begin{acknowledgments}
% Put your acknowledgments here.
%\end{acknowledgments}

% Create the reference section using BibTeX:
%\bibliography{references_bibdesk}

\begin{thebibliography}{0}
\expandafter\ifx\csname natexlab\endcsname\relax\def\natexlab#1{#1}\fi
\expandafter\ifx\csname bibnamefont\endcsname\relax
  \def\bibnamefont#1{#1}\fi
\expandafter\ifx\csname bibfnamefont\endcsname\relax
  \def\bibfnamefont#1{#1}\fi
\expandafter\ifx\csname citenamefont\endcsname\relax
  \def\citenamefont#1{#1}\fi
\expandafter\ifx\csname url\endcsname\relax
  \def\url#1{\texttt{#1}}\fi
\expandafter\ifx\csname urlprefix\endcsname\relax\def\urlprefix{URL }\fi
\providecommand{\bibinfo}[2]{#2}
\providecommand{\eprint}[2][]{\url{#2}}

\end{thebibliography}


\begin{thebibliography}{37}
\expandafter\ifx\csname natexlab\endcsname\relax\def\natexlab#1{#1}\fi
\expandafter\ifx\csname bibnamefont\endcsname\relax
  \def\bibnamefont#1{#1}\fi
\expandafter\ifx\csname bibfnamefont\endcsname\relax
  \def\bibfnamefont#1{#1}\fi
\expandafter\ifx\csname citenamefont\endcsname\relax
  \def\citenamefont#1{#1}\fi
\expandafter\ifx\csname url\endcsname\relax
  \def\url#1{\texttt{#1}}\fi
\expandafter\ifx\csname urlprefix\endcsname\relax\def\urlprefix{URL }\fi
\providecommand{\bibinfo}[2]{#2}
\providecommand{\eprint}[2][]{\url{#2}}

\bibitem[{\citenamefont{Su et~al.}(2008)\citenamefont{Su, Greentree, and
  Hollenberg}}]{2008.OpEx.Hollenberg.NV_cavity}
\bibinfo{author}{\bibfnamefont{C.}~\bibnamefont{Su}},
  \bibinfo{author}{\bibfnamefont{A.~D.} \bibnamefont{Greentree}},
  \bibnamefont{and} \bibinfo{author}{\bibfnamefont{L.~C.~L.}
  \bibnamefont{Hollenberg}}, \bibinfo{journal}{Opt. Express}
  \textbf{\bibinfo{volume}{16}}, \bibinfo{pages}{6240} (\bibinfo{year}{2008}).

\bibitem[{\citenamefont{Santori et~al.}(2010)\citenamefont{Santori, Fattal, and
  Yamamoto}}]{santori_single-photon_2010}
\bibinfo{author}{\bibfnamefont{C.}~\bibnamefont{Santori}},
  \bibinfo{author}{\bibfnamefont{D.}~\bibnamefont{Fattal}}, \bibnamefont{and}
  \bibinfo{author}{\bibfnamefont{Y.}~\bibnamefont{Yamamoto}},
  \emph{\bibinfo{title}{Single-photon devices and applications}}
  (\bibinfo{publisher}{Wiley-{VCH}}, \bibinfo{address}{Weinheim},
  \bibinfo{year}{2010}), ISBN \bibinfo{isbn}{9783527408078 {352740807X}}.

\bibitem[{\citenamefont{Tiecke et~al.}(2014)\citenamefont{Tiecke, Thompson,
  de~Leon, Liu, Vuletic, and Lukin}}]{tiecke_nanophotonic_2014}
\bibinfo{author}{\bibfnamefont{T.~G.} \bibnamefont{Tiecke}},
  \bibinfo{author}{\bibfnamefont{J.~D.} \bibnamefont{Thompson}},
  \bibinfo{author}{\bibfnamefont{N.~P.} \bibnamefont{de~Leon}},
  \bibinfo{author}{\bibfnamefont{L.~R.} \bibnamefont{Liu}},
  \bibinfo{author}{\bibfnamefont{V.}~\bibnamefont{Vuletic}}, \bibnamefont{and}
  \bibinfo{author}{\bibfnamefont{M.~D.} \bibnamefont{Lukin}},
  \bibinfo{journal}{Nature} \textbf{\bibinfo{volume}{508}},
  \bibinfo{pages}{241} (\bibinfo{year}{2014}), ISSN \bibinfo{issn}{0028-0836}.

\bibitem[{\citenamefont{Riedrich-M{\"o}ller
  et~al.}(2011)\citenamefont{Riedrich-M{\"o}ller, Kipfstuhl, Hepp, Neu, Pauly,
  M{\"u}cklich, Baur, Wandt, Wolff, Fischer et~al.}}]{riedrich2011one}
\bibinfo{author}{\bibfnamefont{J.}~\bibnamefont{Riedrich-M{\"o}ller}},
  \bibinfo{author}{\bibfnamefont{L.}~\bibnamefont{Kipfstuhl}},
  \bibinfo{author}{\bibfnamefont{C.}~\bibnamefont{Hepp}},
  \bibinfo{author}{\bibfnamefont{E.}~\bibnamefont{Neu}},
  \bibinfo{author}{\bibfnamefont{C.}~\bibnamefont{Pauly}},
  \bibinfo{author}{\bibfnamefont{F.}~\bibnamefont{M{\"u}cklich}},
  \bibinfo{author}{\bibfnamefont{A.}~\bibnamefont{Baur}},
  \bibinfo{author}{\bibfnamefont{M.}~\bibnamefont{Wandt}},
  \bibinfo{author}{\bibfnamefont{S.}~\bibnamefont{Wolff}},
  \bibinfo{author}{\bibfnamefont{M.}~\bibnamefont{Fischer}},
  \bibnamefont{et~al.}, \bibinfo{journal}{Nature Nanotech.}
  \textbf{\bibinfo{volume}{7}}, \bibinfo{pages}{69} (\bibinfo{year}{2011}).

\bibitem[{\citenamefont{Faraon et~al.}(2012)\citenamefont{Faraon, Santori,
  Huang, Acosta, and Beausoleil}}]{Faraon_coupling_2012}
\bibinfo{author}{\bibfnamefont{A.}~\bibnamefont{Faraon}},
  \bibinfo{author}{\bibfnamefont{C.}~\bibnamefont{Santori}},
  \bibinfo{author}{\bibfnamefont{Z.}~\bibnamefont{Huang}},
  \bibinfo{author}{\bibfnamefont{V.~M.} \bibnamefont{Acosta}},
  \bibnamefont{and} \bibinfo{author}{\bibfnamefont{R.~G.}
  \bibnamefont{Beausoleil}}, \bibinfo{journal}{Phys. Rev. Lett.}
  \textbf{\bibinfo{volume}{109}}, \bibinfo{pages}{033604}
  (\bibinfo{year}{2012}), ISSN \bibinfo{issn}{0031-9007, 1079-7114}.

\bibitem[{\citenamefont{Hausmann et~al.}(2013)\citenamefont{Hausmann, Shields,
  Quan, Chu, de~Leon, Evans, Burek, Zibrov, Markham, Twitchen
  et~al.}}]{hausmann_coupling_2013}
\bibinfo{author}{\bibfnamefont{B.~J.~M.} \bibnamefont{Hausmann}},
  \bibinfo{author}{\bibfnamefont{B.~J.} \bibnamefont{Shields}},
  \bibinfo{author}{\bibfnamefont{Q.}~\bibnamefont{Quan}},
  \bibinfo{author}{\bibfnamefont{Y.}~\bibnamefont{Chu}},
  \bibinfo{author}{\bibfnamefont{N.~P.} \bibnamefont{de~Leon}},
  \bibinfo{author}{\bibfnamefont{R.}~\bibnamefont{Evans}},
  \bibinfo{author}{\bibfnamefont{M.~J.} \bibnamefont{Burek}},
  \bibinfo{author}{\bibfnamefont{a.~S.} \bibnamefont{Zibrov}},
  \bibinfo{author}{\bibfnamefont{M.}~\bibnamefont{Markham}},
  \bibinfo{author}{\bibfnamefont{D.~J.} \bibnamefont{Twitchen}},
  \bibnamefont{et~al.}, \bibinfo{journal}{Nano Lett.}
  \textbf{\bibinfo{volume}{13}}, \bibinfo{pages}{5791} (\bibinfo{year}{2013}).

\bibitem[{\citenamefont{Englund et~al.}(2010)\citenamefont{Englund, Shields,
  Rivoire, Hatami, Vu{\v{c}}kovi{\'c}, Park, and
  Lukin}}]{englund_deterministic_2010}
\bibinfo{author}{\bibfnamefont{D.}~\bibnamefont{Englund}},
  \bibinfo{author}{\bibfnamefont{B.}~\bibnamefont{Shields}},
  \bibinfo{author}{\bibfnamefont{K.}~\bibnamefont{Rivoire}},
  \bibinfo{author}{\bibfnamefont{F.}~\bibnamefont{Hatami}},
  \bibinfo{author}{\bibfnamefont{J.}~\bibnamefont{Vu{\v{c}}kovi{\'c}}},
  \bibinfo{author}{\bibfnamefont{H.}~\bibnamefont{Park}}, \bibnamefont{and}
  \bibinfo{author}{\bibfnamefont{M.~D.} \bibnamefont{Lukin}},
  \bibinfo{journal}{Nano Lett.} \textbf{\bibinfo{volume}{10}},
  \bibinfo{pages}{3922} (\bibinfo{year}{2010}), ISSN \bibinfo{issn}{1530-6984}.

\bibitem[{\citenamefont{Eichenfield et~al.}(2009)\citenamefont{Eichenfield,
  Camacho, Chan, Vahala, and Painter}}]{eichenfield2009picogram}
\bibinfo{author}{\bibfnamefont{M.}~\bibnamefont{Eichenfield}},
  \bibinfo{author}{\bibfnamefont{R.}~\bibnamefont{Camacho}},
  \bibinfo{author}{\bibfnamefont{J.}~\bibnamefont{Chan}},
  \bibinfo{author}{\bibfnamefont{K.~J.} \bibnamefont{Vahala}},
  \bibnamefont{and} \bibinfo{author}{\bibfnamefont{O.}~\bibnamefont{Painter}},
  \bibinfo{journal}{Nature} \textbf{\bibinfo{volume}{459}},
  \bibinfo{pages}{550} (\bibinfo{year}{2009}).

\bibitem[{\citenamefont{Lipson}(2007)}]{lipson_silicon_2007}
\bibinfo{author}{\bibfnamefont{M.}~\bibnamefont{Lipson}},
  \bibinfo{journal}{Nature Photon.} \textbf{\bibinfo{volume}{1}},
  \bibinfo{pages}{18} (\bibinfo{year}{2007}), ISSN \bibinfo{issn}{1749-4885}.

\bibitem[{\citenamefont{Li et~al.}(2013)\citenamefont{Li, Trusheim, Gaathon,
  Kisslinger, Cheng, Lu, Su, Yao, Huang, Bayn et~al.}}]{li2013reactive}
\bibinfo{author}{\bibfnamefont{L.}~\bibnamefont{Li}},
  \bibinfo{author}{\bibfnamefont{M.}~\bibnamefont{Trusheim}},
  \bibinfo{author}{\bibfnamefont{O.}~\bibnamefont{Gaathon}},
  \bibinfo{author}{\bibfnamefont{K.}~\bibnamefont{Kisslinger}},
  \bibinfo{author}{\bibfnamefont{C.-J.} \bibnamefont{Cheng}},
  \bibinfo{author}{\bibfnamefont{M.}~\bibnamefont{Lu}},
  \bibinfo{author}{\bibfnamefont{D.}~\bibnamefont{Su}},
  \bibinfo{author}{\bibfnamefont{X.}~\bibnamefont{Yao}},
  \bibinfo{author}{\bibfnamefont{H.-C.} \bibnamefont{Huang}},
  \bibinfo{author}{\bibfnamefont{I.}~\bibnamefont{Bayn}}, \bibnamefont{et~al.},
  \bibinfo{journal}{J. Vac. Sci. Technol. B} \textbf{\bibinfo{volume}{31}},
  \bibinfo{pages}{06FF01} (\bibinfo{year}{2013}).

\bibitem[{\citenamefont{Batalov et~al.}(2009)\citenamefont{Batalov, Jacques,
  Kaiser, Siyushev, Neumann, Rogers, {McMurtrie}, Manson, Jelezko, and
  Wrachtrup}}]{batalov_low_2009}
\bibinfo{author}{\bibfnamefont{A.}~\bibnamefont{Batalov}},
  \bibinfo{author}{\bibfnamefont{V.}~\bibnamefont{Jacques}},
  \bibinfo{author}{\bibfnamefont{F.}~\bibnamefont{Kaiser}},
  \bibinfo{author}{\bibfnamefont{P.}~\bibnamefont{Siyushev}},
  \bibinfo{author}{\bibfnamefont{P.}~\bibnamefont{Neumann}},
  \bibinfo{author}{\bibfnamefont{L.}~\bibnamefont{Rogers}},
  \bibinfo{author}{\bibfnamefont{R.}~\bibnamefont{{McMurtrie}}},
  \bibinfo{author}{\bibfnamefont{N.}~\bibnamefont{Manson}},
  \bibinfo{author}{\bibfnamefont{F.}~\bibnamefont{Jelezko}}, \bibnamefont{and}
  \bibinfo{author}{\bibfnamefont{J.}~\bibnamefont{Wrachtrup}},
  \bibinfo{journal}{Phys. Rev. Lett.} \textbf{\bibinfo{volume}{102}}
  (\bibinfo{year}{2009}), ISSN \bibinfo{issn}{0031-9007, 1079-7114}.

\bibitem[{\citenamefont{Srinivasan and
  Painter}(2007)}]{srinivasan_optical_2007}
\bibinfo{author}{\bibfnamefont{K.}~\bibnamefont{Srinivasan}} \bibnamefont{and}
  \bibinfo{author}{\bibfnamefont{O.}~\bibnamefont{Painter}},
  \bibinfo{journal}{Appl. Phys. Lett.} \textbf{\bibinfo{volume}{90}},
  \bibinfo{pages}{031114} (\bibinfo{year}{2007}).

\bibitem[{\citenamefont{Zhao et~al.}(2012)\citenamefont{Zhao, Fujiwara, and
  Takeuchi}}]{zhao2012suppression}
\bibinfo{author}{\bibfnamefont{H.-Q.} \bibnamefont{Zhao}},
  \bibinfo{author}{\bibfnamefont{M.}~\bibnamefont{Fujiwara}}, \bibnamefont{and}
  \bibinfo{author}{\bibfnamefont{S.}~\bibnamefont{Takeuchi}},
  \bibinfo{journal}{Opt. Express} \textbf{\bibinfo{volume}{20}},
  \bibinfo{pages}{15628} (\bibinfo{year}{2012}).

\bibitem[{\citenamefont{Doherty et~al.}(2013)\citenamefont{Doherty, Manson,
  Delaney, Jelezko, Wrachtrup, and Hollenberg}}]{doherty2013nitrogen}
\bibinfo{author}{\bibfnamefont{M.~W.} \bibnamefont{Doherty}},
  \bibinfo{author}{\bibfnamefont{N.~B.} \bibnamefont{Manson}},
  \bibinfo{author}{\bibfnamefont{P.}~\bibnamefont{Delaney}},
  \bibinfo{author}{\bibfnamefont{F.}~\bibnamefont{Jelezko}},
  \bibinfo{author}{\bibfnamefont{J.}~\bibnamefont{Wrachtrup}},
  \bibnamefont{and} \bibinfo{author}{\bibfnamefont{L.~C.}
  \bibnamefont{Hollenberg}}, \bibinfo{journal}{Phys. Rep.}
  \textbf{\bibinfo{volume}{528}}, \bibinfo{pages}{1} (\bibinfo{year}{2013}).

\bibitem[{\citenamefont{Hanson et~al.}(2006)\citenamefont{Hanson, Gywat, and
  Awschalom}}]{hanson_room-temperature_2006}
\bibinfo{author}{\bibfnamefont{R.}~\bibnamefont{Hanson}},
  \bibinfo{author}{\bibfnamefont{O.}~\bibnamefont{Gywat}}, \bibnamefont{and}
  \bibinfo{author}{\bibfnamefont{D.~D.} \bibnamefont{Awschalom}},
  \bibinfo{journal}{Phys. Rev. B} \textbf{\bibinfo{volume}{74}},
  \bibinfo{pages}{161203} (\bibinfo{year}{2006}).

\bibitem[{\citenamefont{Hodges et~al.}(2012)\citenamefont{Hodges, Li, Lu, Chen,
  Trusheim, Allegri, Yao, Gaathon, Bakhru, and
  Englund}}]{hodges_long-lived_2012}
\bibinfo{author}{\bibfnamefont{J.~S.} \bibnamefont{Hodges}},
  \bibinfo{author}{\bibfnamefont{L.}~\bibnamefont{Li}},
  \bibinfo{author}{\bibfnamefont{M.}~\bibnamefont{Lu}},
  \bibinfo{author}{\bibfnamefont{E.~H.} \bibnamefont{Chen}},
  \bibinfo{author}{\bibfnamefont{M.~E.} \bibnamefont{Trusheim}},
  \bibinfo{author}{\bibfnamefont{S.}~\bibnamefont{Allegri}},
  \bibinfo{author}{\bibfnamefont{X.}~\bibnamefont{Yao}},
  \bibinfo{author}{\bibfnamefont{O.}~\bibnamefont{Gaathon}},
  \bibinfo{author}{\bibfnamefont{H.}~\bibnamefont{Bakhru}}, \bibnamefont{and}
  \bibinfo{author}{\bibfnamefont{D.}~\bibnamefont{Englund}},
  \bibinfo{journal}{New J. Phys.} \textbf{\bibinfo{volume}{14}},
  \bibinfo{pages}{093004} (\bibinfo{year}{2012}), ISSN
  \bibinfo{issn}{1367-2630}.

\bibitem[{\citenamefont{Carter et~al.}(2013)\citenamefont{Carter, Sweeney, Kim,
  Kim, Solenov, Economou, Reinecke, Yang, Bracker, and
  Gammon}}]{carter2013quantum}
\bibinfo{author}{\bibfnamefont{S.~G.} \bibnamefont{Carter}},
  \bibinfo{author}{\bibfnamefont{T.~M.} \bibnamefont{Sweeney}},
  \bibinfo{author}{\bibfnamefont{M.}~\bibnamefont{Kim}},
  \bibinfo{author}{\bibfnamefont{C.~S.} \bibnamefont{Kim}},
  \bibinfo{author}{\bibfnamefont{D.}~\bibnamefont{Solenov}},
  \bibinfo{author}{\bibfnamefont{S.~E.} \bibnamefont{Economou}},
  \bibinfo{author}{\bibfnamefont{T.~L.} \bibnamefont{Reinecke}},
  \bibinfo{author}{\bibfnamefont{L.}~\bibnamefont{Yang}},
  \bibinfo{author}{\bibfnamefont{A.~S.} \bibnamefont{Bracker}},
  \bibnamefont{and} \bibinfo{author}{\bibfnamefont{D.}~\bibnamefont{Gammon}},
  \bibinfo{journal}{Nature Photon.} \textbf{\bibinfo{volume}{7}},
  \bibinfo{pages}{329} (\bibinfo{year}{2013}).

\bibitem[{\citenamefont{Tartakovskii}(2014)}]{tartakovskii_quantum_2014}
\bibinfo{author}{\bibfnamefont{A.}~\bibnamefont{Tartakovskii}},
  \bibinfo{journal}{Nature Photon.} \textbf{\bibinfo{volume}{8}},
  \bibinfo{pages}{427} (\bibinfo{year}{2014}), ISSN \bibinfo{issn}{1749-4885}.

\bibitem[{\citenamefont{Bernien et~al.}(2013)\citenamefont{Bernien, Hensen,
  Pfaff, Koolstra, Blok, Robledo, Taminiau, Markham, Twitchen, Childress
  et~al.}}]{bernien2013heralded}
\bibinfo{author}{\bibfnamefont{H.}~\bibnamefont{Bernien}},
  \bibinfo{author}{\bibfnamefont{B.}~\bibnamefont{Hensen}},
  \bibinfo{author}{\bibfnamefont{W.}~\bibnamefont{Pfaff}},
  \bibinfo{author}{\bibfnamefont{G.}~\bibnamefont{Koolstra}},
  \bibinfo{author}{\bibfnamefont{M.}~\bibnamefont{Blok}},
  \bibinfo{author}{\bibfnamefont{L.}~\bibnamefont{Robledo}},
  \bibinfo{author}{\bibfnamefont{T.}~\bibnamefont{Taminiau}},
  \bibinfo{author}{\bibfnamefont{M.}~\bibnamefont{Markham}},
  \bibinfo{author}{\bibfnamefont{D.}~\bibnamefont{Twitchen}},
  \bibinfo{author}{\bibfnamefont{L.}~\bibnamefont{Childress}},
  \bibnamefont{et~al.}, \bibinfo{journal}{Nature}
  \textbf{\bibinfo{volume}{497}}, \bibinfo{pages}{86} (\bibinfo{year}{2013}).

\bibitem[{\citenamefont{Pfaff et~al.}(2014)\citenamefont{Pfaff, Hensen,
  Bernien, van Dam, Blok, Taminiau, Tiggelman, Schouten, Markham, Twitchen
  et~al.}}]{pfaff2014unconditional}
\bibinfo{author}{\bibfnamefont{W.}~\bibnamefont{Pfaff}},
  \bibinfo{author}{\bibfnamefont{B.}~\bibnamefont{Hensen}},
  \bibinfo{author}{\bibfnamefont{H.}~\bibnamefont{Bernien}},
  \bibinfo{author}{\bibfnamefont{S.}~\bibnamefont{van Dam}},
  \bibinfo{author}{\bibfnamefont{M.}~\bibnamefont{Blok}},
  \bibinfo{author}{\bibfnamefont{T.}~\bibnamefont{Taminiau}},
  \bibinfo{author}{\bibfnamefont{M.}~\bibnamefont{Tiggelman}},
  \bibinfo{author}{\bibfnamefont{R.}~\bibnamefont{Schouten}},
  \bibinfo{author}{\bibfnamefont{M.}~\bibnamefont{Markham}},
  \bibinfo{author}{\bibfnamefont{D.}~\bibnamefont{Twitchen}},
  \bibnamefont{et~al.}, \bibinfo{journal}{Science}
  \textbf{\bibinfo{volume}{345}}, \bibinfo{pages}{532} (\bibinfo{year}{2014}).

\bibitem[{\citenamefont{Heshami et~al.}(2014)\citenamefont{Heshami, Santori,
  Khanaliloo, Healey, Acosta, Barclay, and Simon}}]{heshami2013raman}
\bibinfo{author}{\bibfnamefont{K.}~\bibnamefont{Heshami}},
  \bibinfo{author}{\bibfnamefont{C.}~\bibnamefont{Santori}},
  \bibinfo{author}{\bibfnamefont{B.}~\bibnamefont{Khanaliloo}},
  \bibinfo{author}{\bibfnamefont{C.}~\bibnamefont{Healey}},
  \bibinfo{author}{\bibfnamefont{V.~M.} \bibnamefont{Acosta}},
  \bibinfo{author}{\bibfnamefont{P.~E.} \bibnamefont{Barclay}},
  \bibnamefont{and} \bibinfo{author}{\bibfnamefont{C.}~\bibnamefont{Simon}},
  \bibinfo{journal}{Phys. Rev. A} \textbf{\bibinfo{volume}{89}},
  \bibinfo{pages}{040301} (\bibinfo{year}{2014}).

\bibitem[{\citenamefont{Childress et~al.}(2006)\citenamefont{Childress, Taylor,
  Sorensen, and Lukin}}]{childress2006fault}
\bibinfo{author}{\bibfnamefont{L.}~\bibnamefont{Childress}},
  \bibinfo{author}{\bibfnamefont{J.}~\bibnamefont{Taylor}},
  \bibinfo{author}{\bibfnamefont{A.~S.} \bibnamefont{Sorensen}},
  \bibnamefont{and} \bibinfo{author}{\bibfnamefont{M.}~\bibnamefont{Lukin}},
  \bibinfo{journal}{Phys. Rev. Lett.} \textbf{\bibinfo{volume}{96}},
  \bibinfo{pages}{070504} (\bibinfo{year}{2006}).

\bibitem[{\citenamefont{Bar-Gill et~al.}(2013)\citenamefont{Bar-Gill, Pham,
  Jarmola, Budker, and Walsworth}}]{bar2013solid}
\bibinfo{author}{\bibfnamefont{N.}~\bibnamefont{Bar-Gill}},
  \bibinfo{author}{\bibfnamefont{L.~M.} \bibnamefont{Pham}},
  \bibinfo{author}{\bibfnamefont{A.}~\bibnamefont{Jarmola}},
  \bibinfo{author}{\bibfnamefont{D.}~\bibnamefont{Budker}}, \bibnamefont{and}
  \bibinfo{author}{\bibfnamefont{R.~L.} \bibnamefont{Walsworth}},
  \bibinfo{journal}{Nat. Commun.} \textbf{\bibinfo{volume}{4}},
  \bibinfo{pages}{1743} (\bibinfo{year}{2013}).

\bibitem[{\citenamefont{Cirac et~al.}(1997)\citenamefont{Cirac, Zoller, Kimble,
  and Mabuchi}}]{CZKM1997PRL}
\bibinfo{author}{\bibfnamefont{J.~I.} \bibnamefont{Cirac}},
  \bibinfo{author}{\bibfnamefont{P.}~\bibnamefont{Zoller}},
  \bibinfo{author}{\bibfnamefont{H.~J.} \bibnamefont{Kimble}},
  \bibnamefont{and} \bibinfo{author}{\bibfnamefont{H.}~\bibnamefont{Mabuchi}},
  \bibinfo{journal}{Phys. Rev. Lett.} \textbf{\bibinfo{volume}{78}},
  \bibinfo{pages}{3221} (\bibinfo{year}{1997}).

\bibitem[{\citenamefont{Kimble}(2008)}]{kimble_quantum_2008}
\bibinfo{author}{\bibfnamefont{H.~J.} \bibnamefont{Kimble}},
  \bibinfo{journal}{Nature} \textbf{\bibinfo{volume}{453}},
  \bibinfo{pages}{1023} (\bibinfo{year}{2008}).

\bibitem[{\citenamefont{Kim et~al.}(2013)\citenamefont{Kim, Bose, Shen,
  Solomon, and Waks}}]{kim2013quantum}
\bibinfo{author}{\bibfnamefont{H.}~\bibnamefont{Kim}},
  \bibinfo{author}{\bibfnamefont{R.}~\bibnamefont{Bose}},
  \bibinfo{author}{\bibfnamefont{T.~C.} \bibnamefont{Shen}},
  \bibinfo{author}{\bibfnamefont{G.~S.} \bibnamefont{Solomon}},
  \bibnamefont{and} \bibinfo{author}{\bibfnamefont{E.}~\bibnamefont{Waks}},
  \bibinfo{journal}{Nature Photon.} \textbf{\bibinfo{volume}{7}}
  (\bibinfo{year}{2013}).

\bibitem[{\citenamefont{Noda et~al.}(2007)\citenamefont{Noda, Fujita, and
  Asano}}]{noda2007spontaneous}
\bibinfo{author}{\bibfnamefont{S.}~\bibnamefont{Noda}},
  \bibinfo{author}{\bibfnamefont{M.}~\bibnamefont{Fujita}}, \bibnamefont{and}
  \bibinfo{author}{\bibfnamefont{T.}~\bibnamefont{Asano}},
  \bibinfo{journal}{Nature Photon.} \textbf{\bibinfo{volume}{1}},
  \bibinfo{pages}{449} (\bibinfo{year}{2007}).

\bibitem[{\citenamefont{Wolters et~al.}(2014)\citenamefont{Wolters, Kabuss,
  Knorr, and Benson}}]{wolters_deterministic_2014}
\bibinfo{author}{\bibfnamefont{J.}~\bibnamefont{Wolters}},
  \bibinfo{author}{\bibfnamefont{J.}~\bibnamefont{Kabuss}},
  \bibinfo{author}{\bibfnamefont{A.}~\bibnamefont{Knorr}}, \bibnamefont{and}
  \bibinfo{author}{\bibfnamefont{O.}~\bibnamefont{Benson}},
  \bibinfo{journal}{arXiv:1405.7307}  (\bibinfo{year}{2014}).

\bibitem[{\citenamefont{Burkard and
  Awschalom}(2014)}]{burkard_cavity-mediated_2014}
\bibinfo{author}{\bibfnamefont{G.}~\bibnamefont{Burkard}} \bibnamefont{and}
  \bibinfo{author}{\bibfnamefont{D.~D.} \bibnamefont{Awschalom}},
  \bibinfo{journal}{arXiv:1402.6351}  (\bibinfo{year}{2014}).

\bibitem[{\citenamefont{Neumann et~al.}(2010)\citenamefont{Neumann, Beck,
  Steiner, Rempp, Fedder, Hemmer, Wrachtrup, and Jelezko}}]{neumann2010single}
\bibinfo{author}{\bibfnamefont{P.}~\bibnamefont{Neumann}},
  \bibinfo{author}{\bibfnamefont{J.}~\bibnamefont{Beck}},
  \bibinfo{author}{\bibfnamefont{M.}~\bibnamefont{Steiner}},
  \bibinfo{author}{\bibfnamefont{F.}~\bibnamefont{Rempp}},
  \bibinfo{author}{\bibfnamefont{H.}~\bibnamefont{Fedder}},
  \bibinfo{author}{\bibfnamefont{P.~R.} \bibnamefont{Hemmer}},
  \bibinfo{author}{\bibfnamefont{J.}~\bibnamefont{Wrachtrup}},
  \bibnamefont{and} \bibinfo{author}{\bibfnamefont{F.}~\bibnamefont{Jelezko}},
  \bibinfo{journal}{Science} \textbf{\bibinfo{volume}{329}},
  \bibinfo{pages}{542} (\bibinfo{year}{2010}).

\bibitem[{\citenamefont{Dolde et~al.}(2013)\citenamefont{Dolde, Jakobi,
  Naydenov, Zhao, Pezzagna, Trautmann, Meijer, Neumann, Jelezko, and
  Wrachtrup}}]{dolde2013room}
\bibinfo{author}{\bibfnamefont{F.}~\bibnamefont{Dolde}},
  \bibinfo{author}{\bibfnamefont{I.}~\bibnamefont{Jakobi}},
  \bibinfo{author}{\bibfnamefont{B.}~\bibnamefont{Naydenov}},
  \bibinfo{author}{\bibfnamefont{N.}~\bibnamefont{Zhao}},
  \bibinfo{author}{\bibfnamefont{S.}~\bibnamefont{Pezzagna}},
  \bibinfo{author}{\bibfnamefont{C.}~\bibnamefont{Trautmann}},
  \bibinfo{author}{\bibfnamefont{J.}~\bibnamefont{Meijer}},
  \bibinfo{author}{\bibfnamefont{P.}~\bibnamefont{Neumann}},
  \bibinfo{author}{\bibfnamefont{F.}~\bibnamefont{Jelezko}}, \bibnamefont{and}
  \bibinfo{author}{\bibfnamefont{J.}~\bibnamefont{Wrachtrup}},
  \bibinfo{journal}{Nature Phys.} \textbf{\bibinfo{volume}{9}},
  \bibinfo{pages}{139} (\bibinfo{year}{2013}).

\bibitem[{\citenamefont{Benson}(2011)}]{benson_assembly_2011}
\bibinfo{author}{\bibfnamefont{O.}~\bibnamefont{Benson}},
  \bibinfo{journal}{Nature} \textbf{\bibinfo{volume}{480}},
  \bibinfo{pages}{193} (\bibinfo{year}{2011}), ISSN \bibinfo{issn}{0028-0836}.

\bibitem[{\citenamefont{Riedrich-M{\"o}ller
  et~al.}(2014)\citenamefont{Riedrich-M{\"o}ller, Arend, Pauly, M{\"u}cklich,
  Fischer, Gsell, Schreck, and Becher}}]{riedrich-moller_deterministic_2014}
\bibinfo{author}{\bibfnamefont{J.}~\bibnamefont{Riedrich-M{\"o}ller}},
  \bibinfo{author}{\bibfnamefont{C.}~\bibnamefont{Arend}},
  \bibinfo{author}{\bibfnamefont{C.}~\bibnamefont{Pauly}},
  \bibinfo{author}{\bibfnamefont{F.}~\bibnamefont{M{\"u}cklich}},
  \bibinfo{author}{\bibfnamefont{M.}~\bibnamefont{Fischer}},
  \bibinfo{author}{\bibfnamefont{S.}~\bibnamefont{Gsell}},
  \bibinfo{author}{\bibfnamefont{M.}~\bibnamefont{Schreck}}, \bibnamefont{and}
  \bibinfo{author}{\bibfnamefont{C.}~\bibnamefont{Becher}},
  \bibinfo{journal}{Nano Lett.} p. \bibinfo{pages}{Article ASAP}
  (\bibinfo{year}{2014}), ISSN \bibinfo{issn}{1530-6984, 1530-6992}.

\bibitem[{\citenamefont{Wolters et~al.}(2013)\citenamefont{Wolters, Sadzak,
  Schell, Schr{\"o}der, and Benson}}]{wolters2013measurement}
\bibinfo{author}{\bibfnamefont{J.}~\bibnamefont{Wolters}},
  \bibinfo{author}{\bibfnamefont{N.}~\bibnamefont{Sadzak}},
  \bibinfo{author}{\bibfnamefont{A.~W.} \bibnamefont{Schell}},
  \bibinfo{author}{\bibfnamefont{T.}~\bibnamefont{Schr{\"o}der}},
  \bibnamefont{and} \bibinfo{author}{\bibfnamefont{O.}~\bibnamefont{Benson}},
  \bibinfo{journal}{Phys. Rev. Lett.} \textbf{\bibinfo{volume}{110}},
  \bibinfo{pages}{027401} (\bibinfo{year}{2013}).

\bibitem[{\citenamefont{Chu et~al.}(2014)\citenamefont{Chu, de~Leon, Shields,
  Hausmann, Evans, Togan, Burek, Markham, Stacey, Zibrov
  et~al.}}]{chu_coherent_2014}
\bibinfo{author}{\bibfnamefont{Y.}~\bibnamefont{Chu}},
  \bibinfo{author}{\bibfnamefont{N.}~\bibnamefont{de~Leon}},
  \bibinfo{author}{\bibfnamefont{B.}~\bibnamefont{Shields}},
  \bibinfo{author}{\bibfnamefont{B.}~\bibnamefont{Hausmann}},
  \bibinfo{author}{\bibfnamefont{R.}~\bibnamefont{Evans}},
  \bibinfo{author}{\bibfnamefont{E.}~\bibnamefont{Togan}},
  \bibinfo{author}{\bibfnamefont{M.~J.} \bibnamefont{Burek}},
  \bibinfo{author}{\bibfnamefont{M.}~\bibnamefont{Markham}},
  \bibinfo{author}{\bibfnamefont{A.}~\bibnamefont{Stacey}},
  \bibinfo{author}{\bibfnamefont{A.}~\bibnamefont{Zibrov}},
  \bibnamefont{et~al.}, \bibinfo{journal}{Nano Lett.}
  \textbf{\bibinfo{volume}{14}}, \bibinfo{pages}{1982} (\bibinfo{year}{2014}),
  ISSN \bibinfo{issn}{1530-6984}.

\bibitem[{\citenamefont{Awschalom et~al.}(2013)\citenamefont{Awschalom,
  Bassett, Dzurak, Hu, and Petta}}]{awschalom_quantum_2013}
\bibinfo{author}{\bibfnamefont{D.~D.} \bibnamefont{Awschalom}},
  \bibinfo{author}{\bibfnamefont{L.~C.} \bibnamefont{Bassett}},
  \bibinfo{author}{\bibfnamefont{A.~S.} \bibnamefont{Dzurak}},
  \bibinfo{author}{\bibfnamefont{E.~L.} \bibnamefont{Hu}}, \bibnamefont{and}
  \bibinfo{author}{\bibfnamefont{J.~R.} \bibnamefont{Petta}},
  \bibinfo{journal}{Science} \textbf{\bibinfo{volume}{339}},
  \bibinfo{pages}{1174} (\bibinfo{year}{2013}), ISSN \bibinfo{issn}{0036-8075,
  1095-9203}.

\bibitem[{\citenamefont{O'Brien et~al.}(2009)\citenamefont{O'Brien, Furusawa,
  and Vu{\v{c}}kovi{\'c}}}]{o2009photonic}
\bibinfo{author}{\bibfnamefont{J.~L.} \bibnamefont{O'Brien}},
  \bibinfo{author}{\bibfnamefont{A.}~\bibnamefont{Furusawa}}, \bibnamefont{and}
  \bibinfo{author}{\bibfnamefont{J.}~\bibnamefont{Vu{\v{c}}kovi{\'c}}},
  \bibinfo{journal}{Nature Photon.} \textbf{\bibinfo{volume}{3}},
  \bibinfo{pages}{687} (\bibinfo{year}{2009}).

\end{thebibliography}

\end{document}